\documentclass[12pt,preprint]{emulateapj}
\usepackage{epsfig}
\usepackage{amsmath}
\usepackage{natbib}

\usepackage{lineno}

\def \teff {$T_{\rm eff}$}
\def \logteff {$\rm log(T_{\rm eff})$}
\def \logg {$\rm log(g)$}
\def \ttau {$T$-$\tau$}

\def \rev {}

\begin{document}


\title{Variation of Stellar Envelope Convection and Overshoot with Metallicity}
\author{Joel D. Tanner, Sarbani Basu \& Pierre Demarque}
\affil{Department of Astronomy, Yale University, P.O. BOX 208101, New Haven, CT 06520-8101}

\begin{abstract}
We examine how metallicity affects convection and overshoot in the superadiabatic layer of main sequence stars.  We present results from a grid of 3D radiation hydrodynamic simulations with four metallicities ($Z=0.040$, $0.020$, $0.010$, $0.001$), and spanning a range in effective temperature ($4950 < \rm T_{\rm eff} < 6230$).  We show that changing the metallicity alters properties of the convective gas dynamics, and the structure of the superadiabatic layer and atmosphere. Our grid of simulations show that the amount of superadiabaticity, which tracks the transition from efficient to inefficient convection, \rev{is sensitive to changes in metallicity.  We find that increasing the metallicity forces the location of the transition region to lower densities and pressures, and results in larger mean and turbulent velocities throughout the superadiabatic region.} We also quantify the degree of convective overshoot in the atmosphere, and show that it increases with metallicity as well.
\end{abstract}

\section{Introduction}

The treatment of convection in stellar envelopes is one of the largest sources of uncertainty in stellar modeling.  Convection in stellar models is usually described by the B\"ohm-Vitense mixing length theory \citep[MLT;][]{1958ZA.....46..108B}, which was derived from Prandtl's mixing hypothesis \citep{prandtl1925}.    MLT represents convection with a single characteristic length, which is proportional to the local pressure scale height $l=\alpha H_P$, where $\alpha$ is a free parameter.  The mixing length parameter is arbitrary, and is usually held at a constant value, obtained from a calibrated solar model. \rev{This single parameter formulation of MLT is commonly used in 1D stellar model calculations, but there are other treatments that introduce additional free parameters to account for geometric properties of convection \citep[see e.g.,][]{2010ApJ...710.1619A}.} While MLT produces an accurate stratification in regions of efficient convection \citep{1989ApJ...336.1022C}, it does not account for the dynamical effects of convection, such as turbulent pressure and asymmetry in the velocity field, and the approximation breaks down near the surface where convection is inefficient.

By fixing the mixing length parameter, the MLT treatment of convection eliminates dependence on stellar properties that are expected to be relevant to convective dynamics.  These properties include the surface gravity and effective temperature of the star, and the chemical composition of the convection zone.  The recent work of \citet{2012ApJ...755L..12B} examined how the mixing length parameter would need to change to satisfy stellar mass and radius constraints from Kepler data. Their results indicate that the mixing length parameter changes with metallicity.  While useful for its application to stellar modeling, the results are limited by the unrealistic MLT representation of convection.  For the past few decades, radiation hydrodynamic simulations (RHD) have been used to provide a more accurate representation of near surface convection.  

Pioneered by \citet{1985pssl.proc..215N}, 3D RHD simulations account for the realistic transition from convective to radiative energy transport.  Since then, simulations have been applied to dwarf stars  \citep[e.g.][]{2009A&A...501.1087R, 2009A&A...508.1429W, 2004IAUS..224..139F}, giants \citep[e.g.][]{2007A&A...469..687C, 2010A&A...524A..93C, 2012A&A...547A.118L}, and several more targeted studies of individual stars \citep[e.g.][]{2005MNRAS.362.1031R, 2006ApJ...636.1078S, 2007IAUS..239..388S,  2009A&A...506..167L, 2010A&A...513A..72B} 

Efforts to systematically study the variation of stellar convection have been carried out by \citet{1995LIACo..32..213L, 1998IAUS..185..115L, 1999A&A...346..111L}, \citet{1999ASPC..173..225F}, and \citet{2011ApJ...731...78T}, and there are several ambitious efforts to examine metallicity and convection using grids of simulations by including a metallicity dimension in the grid parameter space.  These include the ongoing CIFIST \citep{2009MmSAI..80..711L} and STAGGERGRID \citep{2011JPhCS.328a2003C} projects.  Our work follows in a similar vein and employs a grid of 3D RHD simulations to isolate the effect of varying metallicity while fixing other dimensions of parameter space.  

Determining how realistic stellar surface convection is sensitive to metallicity remains an area of active research.  Some of the simulations mentioned above were carried out at different metallicities and compositions, and there are a few studies comparing simulations at varied metallicity.  For example, \citet{2007ASPC..362..306J} examined 3D RHD simulations of convection in the sun and a population II star.  They find differences in the turbulent pressure and kinetic energy, but the stellar parameters corresponding to the simulations have different metallicity, effective temperature, and surface gravity. Similarly, \citet{2009IAUS..259..233S} compared 2D simulations of magneto-convection for the Sun and a metal poor solar-like analog, but with the same surface gravity and effective temperatures.  Their preliminary results show that the primary effect of the lower opacity in the metal-poor star is increased pressure and magnetic field strength.   We aim to perform a more systematic and rigorous study by using a grid of 3D  simulations to isolate the effect of metallicity over a range in effective temperature.  

Other studies have focused on elemental abundances and spectral features.  For example,  \citet{1999A&A...346L..17A} used convection simulations corresponding to two halo stars in conjuction with spectral synthesis to estimate chemical abundances, which indicated that 1D models have led to overestimated primordial Li abundances.  Similarly \citet{2007A&A...469..687C} found differences in the predicted line strengths from 3D simulations of metal-poor red giant stars compared to corresponding 1D model atmospheres.   Their simulations cover a range in effective temperature and metallicity, and they found that spectral lines of neutral metals and molecules appear stronger in 3D simulations than in 1D models.  Results of such studies consistently show cooler atmospheres in lower metallicity simulations.  A recent study by \citet{2011A&A...528A..32C} examined 3D simulations of low-metallicity red giants, and confirmed that the cooler atmospheres in low-Z simulations was not the result of the approximations to the treatment of scattering. 

Most previous studies of the effect of metallicity on convection have focused on atmospheric properties.  We aim to examine how metallicity determines the structure of the convection zone, particularly in the superadiabatic layer where energy transport transitions from convective to radiative, with the intent of improving stellar models.  In the present study, we endeavor to systematically explore the effect of varying heavy metal abundance and convection in the superadiabatic layer by computing several sets of simulations at various metallicities, spanning a range in effective temperature.  The effective temperature range for each metallicity set overlaps, enabling the comparison of simulations at varied $Z$ but fixed {\logg} and {\logteff}.   Section \ref{sec:code} outlines the numerical method and details of the simulation code, while Section \ref{sec:grid} describes the individual simulations that we present.  In Sections \ref{sec:salstructure} through \ref{sec:ttau} we present the effect of metallicity on the thermal structure and gas dynamics.

\section{The Radiation Hydrodynamics Code} \label{sec:code}

Our simulation code solves the compressible Navier-Stokes equations with radiative transfer.  Details are provided in \citet{2012ApJ...759..120T}.  It is an updated version of the code described in \citet{2003MNRAS.340..923R, 2004MNRAS.347.1208R}, and is based on the code of Kim, Chan \& Sofia \citep{1989ApJ...336.1022C, 1998ApJ...496L.121K}.  

The simulation domain is a Cartesian box with periodic boundary conditions at vertical walls and closed surfaces at the top and bottom.  \rev{Our experiments with different boundary conditions consistently show that the effect of the boundary layer is less than one scale height from the edge of the computational domain, and this boundary layer has been removed from the figures.}  The computational domain spans the superadiabatic layer, which is the region of transition from convection to radiation.  The bottom of the domain is fully convective, while the top is fully radiative. Thus, the computational domain spans the transition region from convective to radiative energy transport, where convection is inefficient and MLT is insufficient.

The extent of the simulation domain is approximately 9 pressure scale heights (typically less than 0.1\% of the stellar radius for the simulations presented here), which is very small relative to the full extent of the convection zone (which is approximately 20 pressure scale heights in the sun), so curvature and radial variation of gravity are negligible and can be ignored.  Although the simulation domain is small relative to the full extent of the convection zone, it encompasses precisely the region where MLT fails to capture the effect realistic convection, such as the additional support of turbulent pressure.

Radiative transfer is treated with the diffusion approximation in the optically thick part of the domain.  In the optically thin layers where the diffusion approximation is invalid, the code can solve for the mean intensity using the 3D Eddington approximation \citep{1966PASJ...18...85U} or integration over long ray characteristics (similar to the method described by \citet{2003ASPC..288..519S}.  The simulations were computed using the 3D Eddington solver, which compares well with long characteristic ray integration, and offers some computational performance benefits.  A detailed comparison of the radiative transfer solvers has been presented in \citet{2012ApJ...759..120T}.  The code uses the OPAL equation of state and opacity tables \citep{2002ApJ...576.1064R} and the \citet{2005ApJ...623..585F} opacity tables at low temperatures.

\section{Grid of Simulations} \label{sec:grid}

Each simulation is characterized by its energy flux, surface gravity, and composition.  It is important to note that while we directly set the surface gravity and composition, the energy flux is a quantity that is calculated.  We define the effective temperature of the simulation as  
\begin{equation}
T_{\rm eff}^4 = \frac{F_{\rm rad}}{\sigma}
\end{equation}
where $\sigma$ is Stefan-Boltzmann constant, and the radiative flux, $F_{\rm rad}$, is determined by the radiative transfer solver. Changing the metallicity in a simulation alters the thermal balance in the atmosphere, resulting in a different energy flux.  Without precise control over the radiative energy flux, it is difficult to compare simulations at constant effective temperature and varied metallicity.  To overcome this, we have computed a grid of simulations for several metallicities that span a small range in effective temperature.    

The grid comprises four sets of simulations, each at a different metallicity. Each set spans a small range in effective temperature, which overlap with each other. All simulations in the grid have the same surface gravity set to $\log⁡(g)=4.30$.  Properties of the grid are listed in Table \ref{tab:grid}. We can isolate the effect of metallicity by interpolating within the grid to achieve a desired effective temperature, or by comparing simulations that happen to have similar energy fluxes.  Simulations use opacity and equation of state tables with the heavy element abundance mixture of \citet{1998SSRv...85..161G}.

\rev{Since the computational domain spans a sufficiently small fraction of the stellar radius so that the gravitational field is homogeneous and curvature is neglected, the simulations do not contain information about the stellar radius.  As a result, the zero point of the radial coordinate is arbitrary, and the simulation domain cannot be expressed as a fractional stellar radius.}  Instead of using radius, we typically present properties of the simulations relative to the effective temperature surface.  This reference point (where $\langle{T}\rangle =T_{\rm eff}$) can be interpreted as the stellar `surface' because it is the point at which a black body yields the same energy flux as that of the simulation.  We define the height such that the zero point is fixed at the {\teff} surface, and it increases toward smaller optical depth.

In the following sections we compare space- and time- averaged quantities over the vertical (radial) coordinate.  The horizontal and temporal mean over a time $\Delta t$ and a box cross-sectional area of $L_x \times L_y$ is:
\begin{equation}
\label{eqn:mean}
\langle q \rangle = \frac{1}{\Delta t}\frac{1}{L_x L_y} \int_{t_o}^{t_o+\Delta t} \int_{0}^{L_x} \int_{0}^{L_y} q dx dy dt .
\end{equation}

Note that quantities are averaged over a sufficient interval to obtain statistical convergence, and the start of the time interval is after the simulation has thermally relaxed.  The rms of the same quantity is:
\begin{equation}
\label{eqn:rms}
q_{\rm rms} = \sqrt{\langle{q^2}\rangle - \langle{q}\rangle^2} .
\end{equation}

The correlation function between two turbulent quantities $q_1$ and $q_2$ is defined as:
\begin{equation}
\label{eqn:corr}
C[q_1,q_2] = \frac{\langle{q_1 q_2}\rangle - \langle{q_1}\rangle \langle{q_2}\rangle}{q_{\rm 1,rms}q_{\rm 2,rms}} .
\end{equation}

\begin{table}
  \caption{Properties of the simulations in the grid.  All simulations have the same surface gravity ($\log g = 4.30$).}
  \label{tab:grid}
  \begin{center}
    \leavevmode
    \begin{tabular}{lllll} \hline \hline              
  $ID$          & Z              & X & $T_{\rm eff}$ & $\log T_{\rm eff}$      \\ \hline 
s1 & 	0.040 & 	0.715 & 	4947	&	3.694  \\
s2 & 	0.040 & 	0.715 & 	5201	&	3.716  \\
s3 & 	0.040 & 	0.715 & 	5461	&	3.737  \\
s4 & 	0.040 & 	0.715 & 	5719	&	3.757  \\
s5 & 	0.020 & 	0.735 & 	5121	&	3.709  \\
s6 & 	0.020 & 	0.735 & 	5370	&	3.730  \\
s7 & 	0.020 & 	0.735 & 	5626	&	3.750  \\
s8 & 	0.020 & 	0.735 & 	5897	&	3.770  \\
s9 & 	0.010 & 	0.745 & 	5326	&	3.726  \\
s10 & 	0.010 & 	0.745 & 	5569	&	3.746  \\
s11 & 	0.010 & 	0.745 & 	5803	&	3.764  \\
s12 & 	0.010 & 	0.745 & 	6027	&	3.780  \\
s13 & 	0.001 & 	0.754 & 	5735	&	3.759  \\
s14 & 	0.001 & 	0.754 & 	5896	&	3.771  \\
s15 & 	0.001 & 	0.754 & 	6063	&	3.783  \\
s16 & 	0.001 & 	0.754 & 	6235	&	3.795  \\ \hline
    \end{tabular}
  \end{center}
\end{table}

\section{Results: SAL Structure} \label{sec:salstructure}

\subsection{Mean Stratification} \label{sec:meanstratifications}

\begin{figure}[h]
\epsscale{1.2}
\plotone{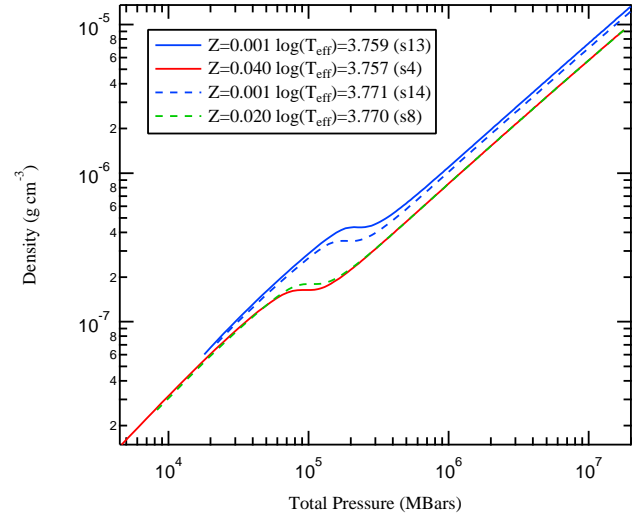}
\caption{Structures through the SAL for four select simulations.  Simulation pair (s13,s14) shows how the stratification changes with effective temperature, while simulations pairs (s4,s13) and (s8,s14) have essentially the same effective temperatures but different metallicities.  The density and pressure of the SAL are sensitive to changes in metallicity.}
\label{fig:rhovsp}
\end{figure}

The structure of the near surface layers is determined largely by the inefficient convection.  It is in precisely this region that convection models such as MLT fail to provide accurate stratifications. Figure \ref{fig:rhovsp}  shows the density as a function of total pressure (which includes both gas and turbulent pressure) for four simulations.  

The simulations in Figure \ref{fig:rhovsp} were selected to show the effect of increasing the energy flux and changing the metallicity. The region of transition from convective to radiative energy transport, known as the superadiabatic layer (SAL), occurs just below the stellar surface. In Figure \ref{fig:rhovsp} the SAL is apparent as an abrupt change in the density-pressure relationship.  Over this pressure range, increasing the effective temperature primarily changes the density in these layers, while changing metallicity substantially changes both the density and the location of the SAL.  Higher metallicity simulations have SALs that occur at lower pressure and density.  

The density in the SAL reflects the change in opacity as a result of changing metallicity. At these temperatures, the primary opacity source is from the $H^-$ ion.  \rev{Increasing the metallicity results in low ionization elements donating more electrons to hydrogen atoms, thereby increasing the opacity.}  The energy flux is determined by $\rho \kappa$ and the source function, so to maintain the same energy flux at higher opacity ($\kappa$), the density ($\rho$) must be correspondingly lower.  

\subsection{Superadiabatic Excess} \label{sec:sal}

\rev{Near the surface, where convection is inefficient, a useful quantity to examine is the radial variation of the superadiabatic excess ($\nabla - \nabla_{\rm ad}$), which is the difference between the temperature gradient, defined as:}
\begin{equation}\label{eqn:sae}
\nabla = \frac{d\ln(T)}{d\ln(P)} ,
\end{equation}
\rev{and the adiabatic temperature gradient.  Turbulent gas dynamics in the SAL contribute a significant fraction to the pressure term in Equation \ref{eqn:sae}.  The total pressure, $P = P_{\rm gas} + P_{\rm turb}$,  is the sum of the gas and turbulent pressures, where the turbulent pressure is calculated with the defintion in Section \ref{sec:pturb}.}  In the deep layers where the structure is nearly isentropic and convection is efficient, the temperature gradient is essentially adiabatic.  Nearer to the stellar surface, where the energy flux transitions from convective to radiative, we see a departure from adiabaticity characterized by a sharp peak in the superadiabatic gradient.  The difference between the two gradients (i.e.\ the superadiabatic excess), provides a measure of the inefficiency of convection, and reveals the depth at which there is a significant departure from efficient convection.

\begin{figure*}
\epsscale{1.0}
\plotone{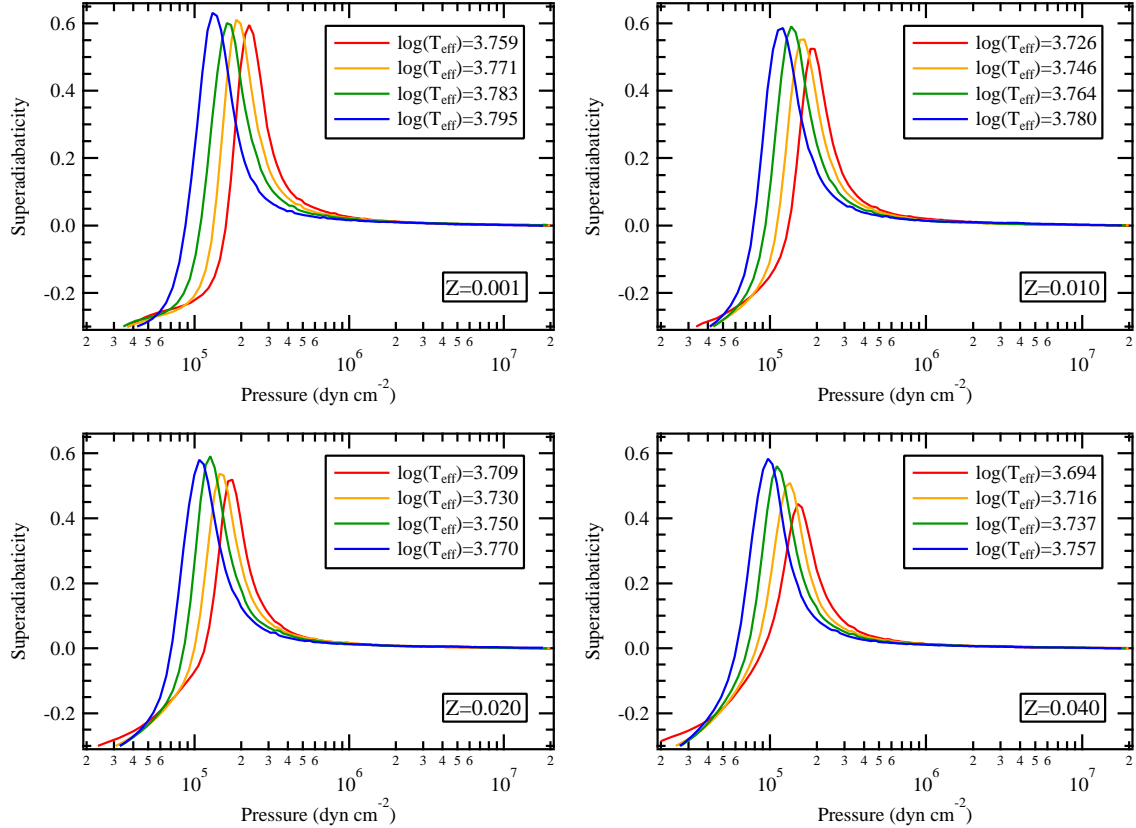}
\caption{Superadiabatic gradients ($\nabla - \nabla_{\rm ad}$) for the grid of simulations.  Each panel shows a set of simulations over a small range in effective temperature and at a given metallicity.  At a given metallicity, the maximum superadiabaticity changes with {\teff}, but at lower metallicity the maximum superadiabaticity becomes less sensitive to {\teff}.}
\label{fig:salpanel}
\end{figure*}

Figure \ref{fig:salpanel} shows the superadiabatic gradients as a function of pressure for our grid of simulations. Each panel in the figure corresponds to a different metallicity set. The temperature structures show a mild systematic change in the maximum suberadiabaticity (maximum of $\nabla - \nabla_{\rm ad}$) with the effective temperature of the simulations, \rev{although the variation is markedly less apparent than what is predicted by stellar models computed with MLT.  The more constrained range in superadiabaticity relative to MLT is generally consistent with the grid of simulations of \citet{2010Ap&SS.328..213T}, which show a smaller variation in the superadiabatic peak compared with MLT models at evolutionary stages from the main sequence to giants.} The simulations with larger effective temperatures have higher SALs, particularly for high-Z simulations, but the maximum suberadiabaticity becomes less sensitive to {\teff} at near-Solar and lower-Z.  We note, however, that while the temperature ranges overlap, they are not identical for each metallicity simulation set.

\begin{figure}[!h]
\epsscale{1.2}
\plotone{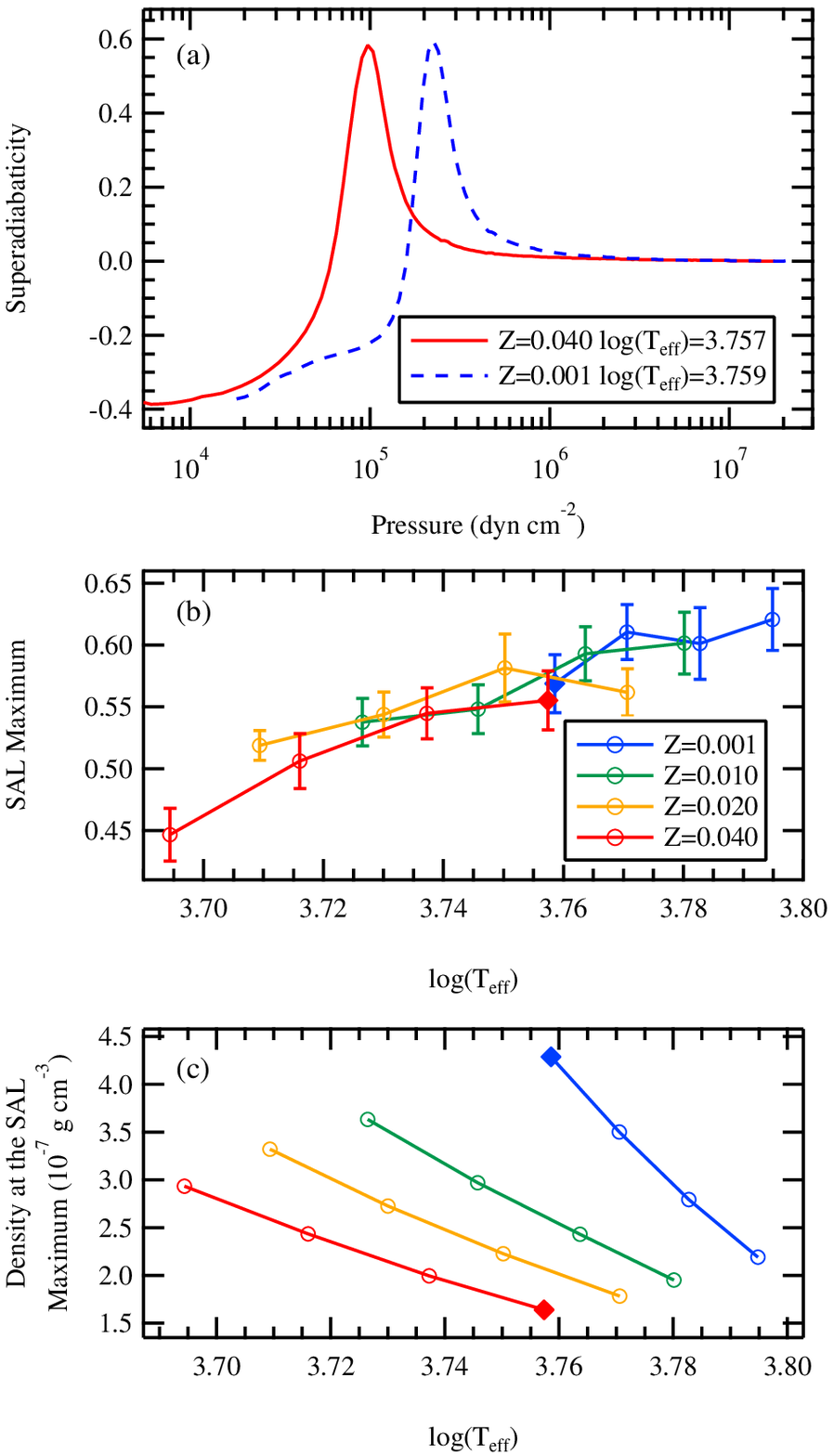}
\caption{(\textit{a}): Superadiabatic gradients from simulations s4 and s13, which have different metallicities but the same effective temperature.  The maximum of the SAL remains essentially unchanged despite the considerable difference in metallicity.  The location of the SAL is sensitive to metallicity.  (\textit{b}): The maximum value of the superadiabaticity and (\textit{c}) location of the maximum superadiabaticity for all simulations in the grid.  Each line in panels (\textit{b}) and (\textit{c}) corresponds to a set of simulations with a fixed metallicity.  Simulations s4 and s13 are marked with diamond symbols.  Metallicity does not appear to directly affect the maximum of the SAL, but it does introduce a shift in the location of the SAL with respect to density.}
\label{fig:saltrends}
\end{figure}

When the metallicity is changed and {\teff} is fixed, we find no significant change in the maximum superadiabaticity.  The top panel of Figure \ref{fig:saltrends} compares the superadiabatic gradients in two simulations that have almost the same effective temperatures (the effective temperatures differ by only $27K$) but a drastic change in metallicity (Z differs by a factor of 40).  In both simulations the maximum superadiabatic excess is approximately the same.  We do, however, see a marked shift in the depth of the SAL.  At lower metallicity, the conversion from efficient to inefficient convection happens at higher pressures and densities. 

The degree of superadiabaticity corresponds to the rate at which energy transport transitions from convective to radiative.  A large value of the maximum SAL is the result of a steep entropy gradient.  The relatively constant SAL maximum as a function of metallicity suggests that the thermal structure changes (examined in section \ref{sec:meanstratifications}), rather than a large change in the convection-to-radiation transition.  

We summarize the simulated SALs in the lower panels of Figure \ref{fig:saltrends}, which show the maximum superadiabaticity and its location for all the simulations in the grid.  In the top panel, each metallicity set shows a shallow trend with effective temperature, \rev{but there is no remarkable offset between them.} Although the maximum value of the SAL remains unchanged, the bottom panel of Figure \ref{fig:saltrends} shows that \rev{the change in} metallicity introduces a shift in the location of the SAL.  Lower-Z simulations have correspondingly higher densities to maintain the same energy flux.  Thus, radiative energy transport starts to become significant at higher density in the lower-Z simulations.  

\section{Results: Dynamics of Convection} \label{sec:dynamics}

\subsection{Mean Convective Velocities} \label{sec:vel}

\begin{figure*}
\begin{minipage}[b]{0.9\linewidth}
\epsscale{1.0}
\plotone{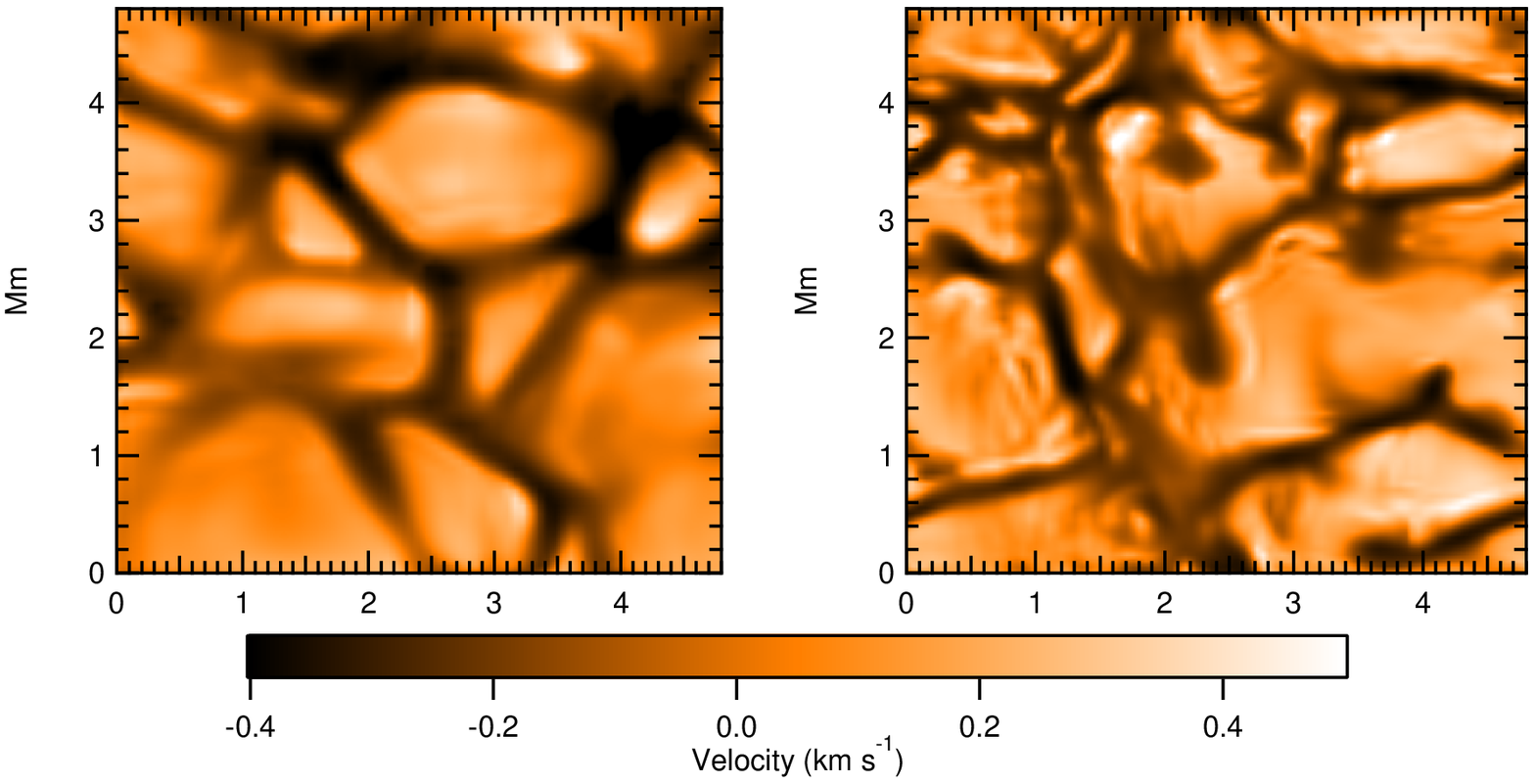}
\caption{Vertical velocity at the effective temperature surface showing the characteristic granulation pattern from snapshots from two simulations in the grid. The two simulations have the same surface gravity and effective temperature, but metallicities of $Z=0.040$ (s4, left) and $Z=0.001$ (s13, right).}
\label{fig:granulation}
\end{minipage}
\end{figure*}

One of the distinctions between mixing length theory and realistic convection is the asymmetry of the velocity field \rev{resulting from upflows being characterized by a larger filling factor at almost all depths in the simulation domain}.  The hot upflowing gas rises nearly adiabatically at approximately constant entropy \citep[e.g.][]{1999A&A...346..111L,1999AAS...194.2104S} until it radiates away energy near the surface before cycling back into the convection zone.  Stellar convection is characterized by large upflowing regions (granules) separated by cool fast downdrafts (intergranular lanes).  Figure \ref{fig:granulation} shows the vertical velocity at a horizontal slice where $\langle T \rangle =T_{\rm eff}$ for two simulations with different metallicities.  The flow pattern qualitatively resembles solar granulation with large expanding updrafts separated by cool fast downdrafts.  The difference in metallicity, however, causes the granulation pattern to be visibly different.  Granulation in the higher-Z simulation shows larger granules and sharper inter-granular lanes.

\begin{figure*}
\begin{minipage}[b]{0.9\linewidth}
\epsscale{1.2}
\plotone{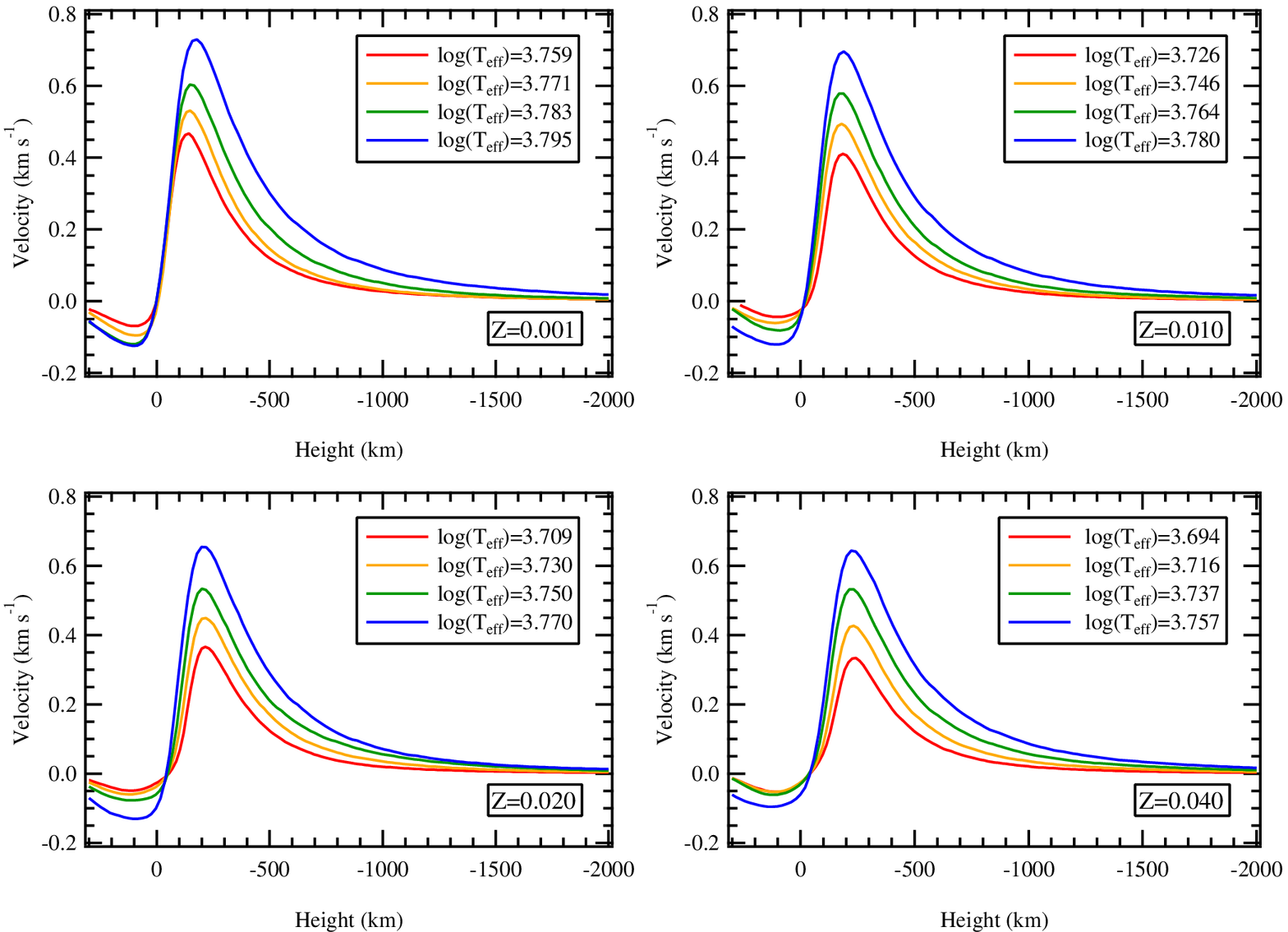}
\caption{Radial variation of average vertical velocities for all simulations in the grid.  Velocities are small in the deep region, and reach a maximum near the peak of the superadiabaticity.  Larger velocities are measured in simulations with higher metallicity or effective temperature.}
\label{fig:wvelpanel}
\end{minipage}
\end{figure*}

Asymmetry in the velocity field means that the average vertical velocity, $\langle w \rangle$, is dominated by the upflows through most of the simulation domain.  As shown in Figure \ref{fig:wvelpanel}, the average vertical velocities are small deep in the convection zone and rise to a maximum just below the surface near the peak of the SAL. The velocity profile is inverted above the convectively stable region (where $\nabla < \nabla_{\rm ad}$), and the average velocity becomes negative.   

\begin{figure}
\epsscale{1.2}
\plotone{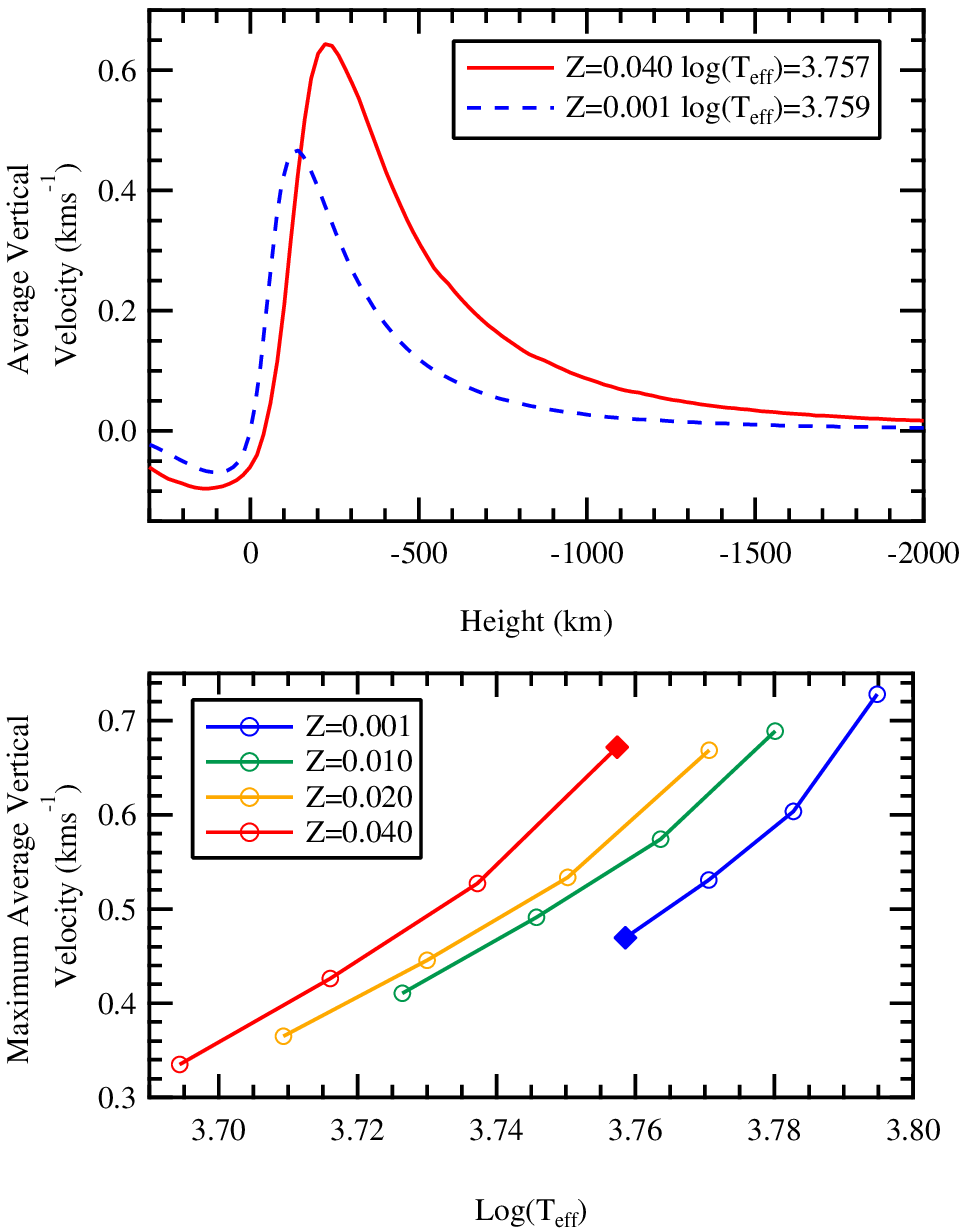}
\caption{(\textit{top}): Radial variation of vertical velocity for simulations s4 and s13 (identified with diamond symbols in the bottom panel), which have approximately the same {\teff}.   (\textit{bottom}): The maximum average vertical velocity for all simulations in the grid.  The maximum velocity increases with effective temperature and metallicity.}
\label{fig:wveltrends}
\end{figure}

The maximum average vertical velocities of the different simulations are compared in Figure \ref{fig:wveltrends}. The trend with effective temperature (from velocity profiles in Figure \ref{fig:wvelpanel}) is apparent, but there is also a distinct metallicity dependence. At a given effective temperature, simulations with higher metallicity exhibit larger convective velocities.

The trend is a direct result of the changes in density stratification, which change in response to the different opacities caused by changing the metallicity (see Section \ref{sec:meanstratifications}).  In simulations with the same {\teff}, convection is responsible for transferring the same energy flux.  The different stratifications, however, require different velocities.  High-Z simulations have higher opacity, and consequently lower density through the SAL (see Figure \ref{fig:rhovsp}), so the convective velocities must be larger to sustain the same convective flux.

Larger convective velocities in the SAL mean that upflows will remain coherent over a larger spatial scale, even in the radiative domain where convection is no longer driven.  In Section \ref{sec:overshoot} we examine the strength of convection beyond the convectively stable boundary, and provide a measure of how it changes with metallicity.

\subsection{Turbulent Pressure} \label{sec:pturb}

One of the physical phenomena absent in MLT-like prescriptions for convection is the additional pressure support from turbulence.  The so-called turbulent pressure is caused by turbulent fluctuations in the velocity field, and is a non-negligible contribution to hydrostatic balance through the SAL, and causes structural differences where turbulent pressure is significant.

We estimate turbulent pressure from the rms (Equation \ref{eqn:rms}) of the vertical velocity field as:
\begin{equation}
P_{\rm turb} = \rho w_{\rm rms} .
\end{equation}

Simulations of solar convection estimate turbulent pressure support at approximately 15\% to 18\% of the gas pressure near the peak of the SAL \citep{2009MmSAI..80..701K, 2012A&A...539A.121B}. We see similar results from our code. Turbulent pressure is negligible deep in the near-adiabatic part of the convection zone because of extremely large gas pressure, and is also quite small in the optically thin region where the gas density and rms velocities are small.

\begin{figure*}
\begin{minipage}[b]{0.9\linewidth}
\epsscale{1.2}
\plotone{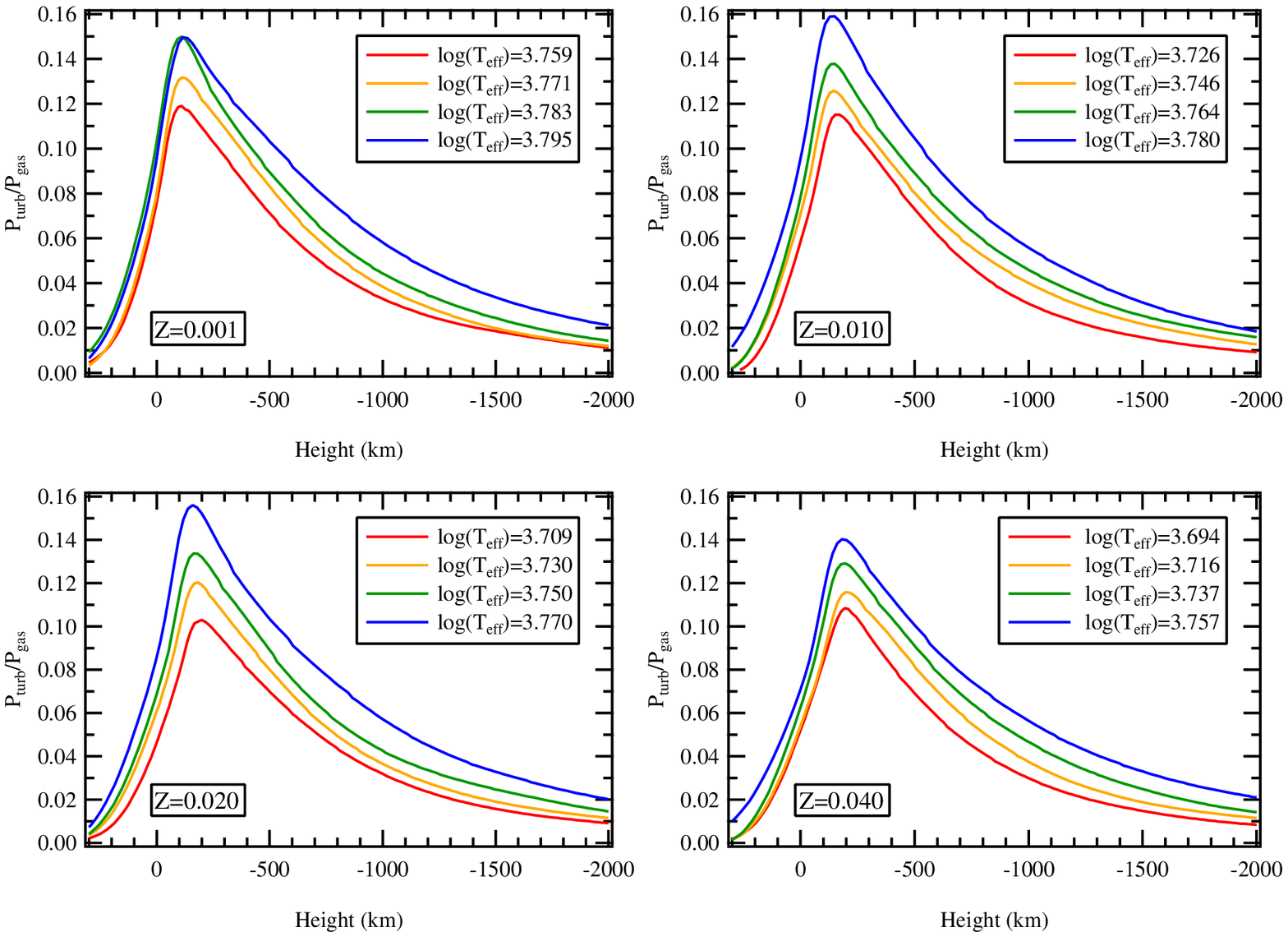}
\caption{Radial variation of average turbulent pressure for all simulations in the grid.  Turbulent pressure follows the same trend as the mean velocities (Figure \ref{fig:wvelpanel}), increasing with effective temperature and metallicity.}
\label{fig:pturbpanel}
\end{minipage}
\end{figure*}

Figure \ref{fig:pturbpanel} compares the radial variation of turbulent pressure (relative to gas pressure) from all of the simulations. Turbulent pressure is a few percent in the deep convection zone, and increases at smaller depth, reaching a maximum near the peak of the SAL. Turbulent pressures in our grid can be as high as 10-16\% of the gas pressure.

\begin{figure}[!h]
\epsscale{1.2}
\plotone{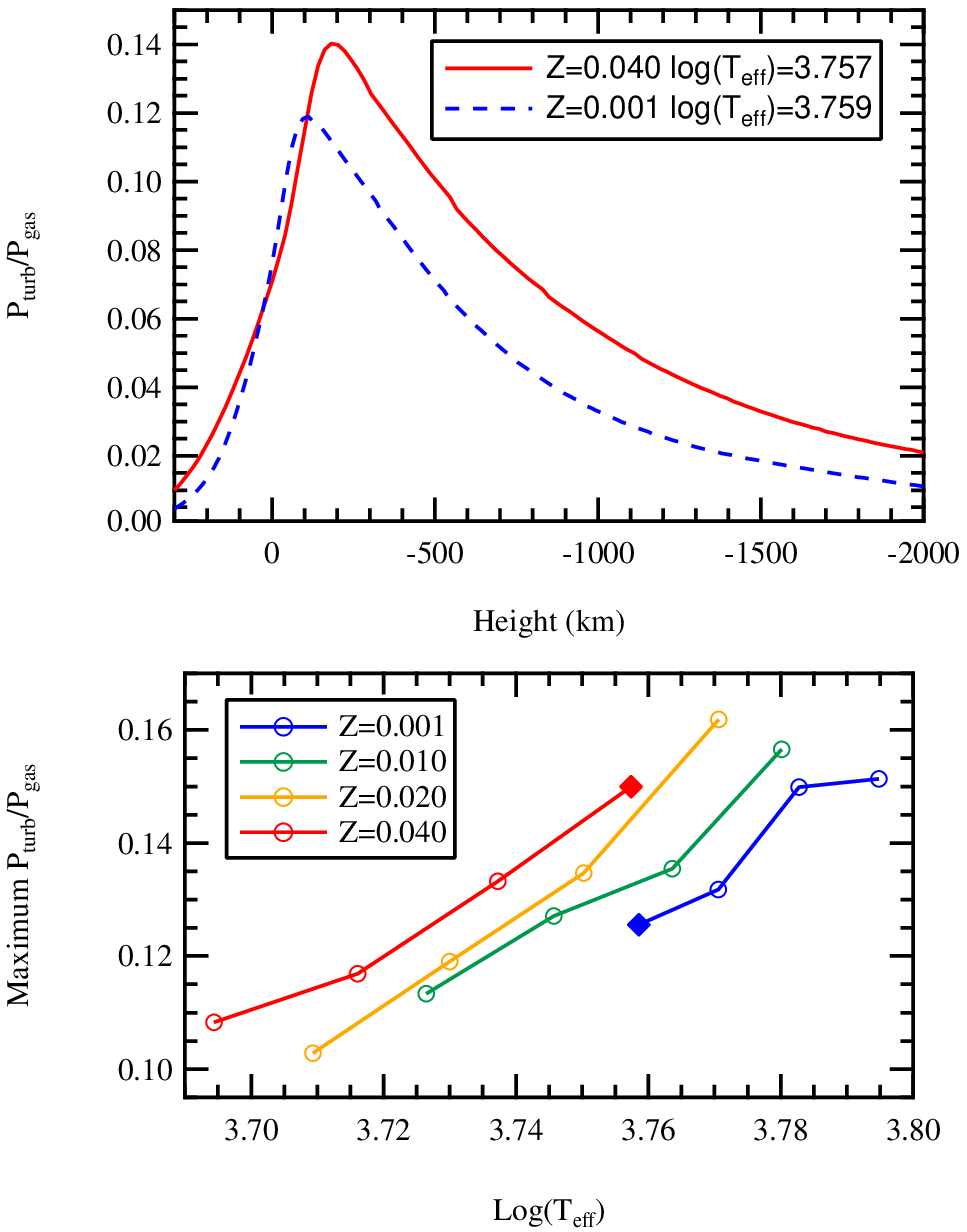}
\caption{(\textit{top}):  Radial variation of turbulent pressure for simulations s4 and s13 (identified with diamond symbols in the bottom panel), which have approximately the same {\teff}. (\textit{bottom}):  Maximum average turbulent pressure for all simulations in the grid.  Simulations with higher effective temperatures or metallicities show increased turbulent pressure.}
\label{fig:pturbtrends}
\end{figure}

As with the mean vertical velocities, we find that the turbulent pressure support scales with both the energy flux and metallicity. The high-Z simulations show more support from turbulence \rev{(as a fraction of gas pressure)} relative to the low-Z simulations. Comparing the turbulent pressure from each simulation (Figure \ref{fig:pturbtrends}) shows that the maximum turbulent pressure visibly increases with effective temperature, and there is a clear systematic offset that is correlated with changes in metallicity.  The increase in turbulent pressure is a result of larger rms velocities, which are also correlated with effective temperature and metallicity.

\subsection{Overshoot} \label{sec:overshoot}

Convective regions in 1D stellar models are well defined and unambiguous.  The radial domain that is unstable to convection can be defined according to many equivalent criteria, such as $\nabla_{\rm ad} < \nabla$, $ds/dz<0$, or $F_{\rm conv}>0$.  Stratifications from 3D RHD simulations do not provide such clear boundaries, and there are several ways to estimate the edge of a driven convection zone, or the extent of a region influenced by convection. 

It is well known that updrafts would not simply stop at the edge of the convectively unstable layer.  Instead, their momentum would carry them some distance beyond the convective boundary effectively extending the region of mixing.  The amount of additional mixing, or overshoot, will depend on the stratification of the atmosphere and the convective dynamics.  

Simulations treat convection and the effect of overshoot self-consistently, and make no distinction between them. We can, however, quantify the amount of ‘overshoot’ by measuring correlations between turbulent fluctuations.  In convectively unstable regions, the upflows are coherent and turbulent fluctuations are highly correlated.  By measuring the vertical (radial) location where the fluctuations are no longer correlated, we can determine the height at which the convection no longer resembles the behavior of the unstable region.

We define the end of the convectively unstable region (and the start of the ‘overshoot’ region) to be the height at which $\nabla-\nabla_{\rm ad}$ is zero (see Figure \ref{fig:salpanel}).  We define the end of the ‘overshoot’ region to be where fluctuations in temperature and vertical velocity are no longer correlated, where the degree of correlation is defined by Equation \ref{eqn:corr}.  This definition has been used by \citet{1989ApJ...336.1022C} and \citet{2004MNRAS.347.1208R}. The size of the ‘overshoot’ region is the difference between these two heights.

\begin{figure*}
\begin{minipage}[b]{0.9\linewidth}
\epsscale{1.2}
\plotone{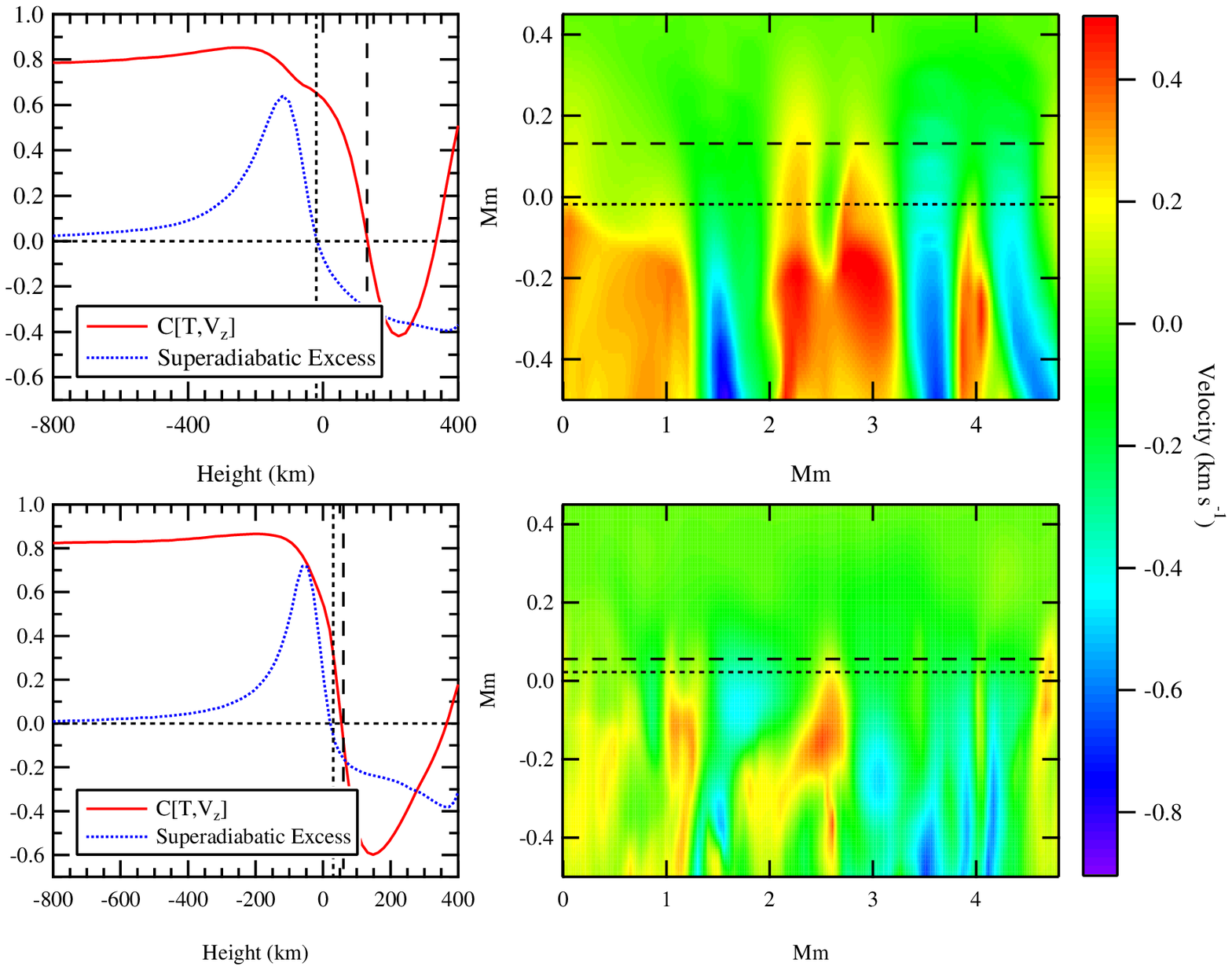}
\caption{Convective overshoot for simulations s4 (top) and s13 (bottom).  The left side shows the superadiabatic excess and correlation functions that are used to define the region of overshoot.  The right side shows the slices of the vertical velocity near the {\teff} surface.  Dashed lines show the measured boundaries of the overshoot region.  The high-Z simulation (s4) has larger velocities and more overshoot.}
\label{fig:overshoot}
\end{minipage}
\end{figure*}

The correlation through the deeper part of the convection zone is quite constant over this range in {\teff} and Z, ranging from  $C[T,V_z] = 0.77-0.85$, which is similar to the value measured by Chan \& Sofia (1989).  The rate at which the correlation drops, however, is sensitive to metallicity.  Figure \ref{fig:overshoot} compares the temperature-velocity correlations and superadiabatic gradients in two simulations with different metallicity and the same effective temperature.  The zero point of the radial height is defined to match the average effective temperature surface. The edge of the convectively unstable region, as defined by where the superadiabatic gradient is zero, is almost unchanged in the two simulations. The region affected by overshooting, as measured by the temperature-velocity correlation, is clearly larger in the higher metallicity simulation.

\begin{figure}
\epsscale{1.2}
\plotone{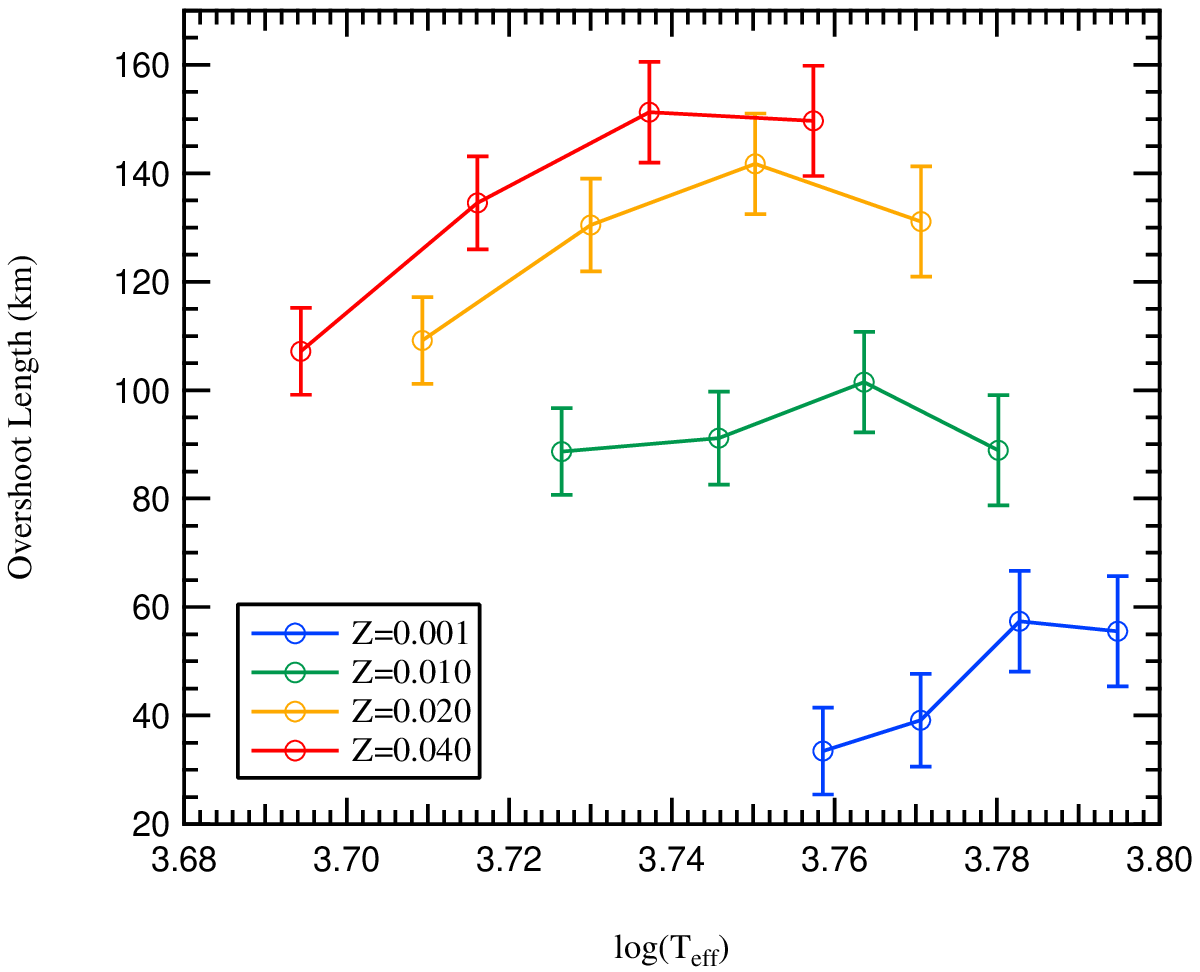}
\caption{The extent of convective overshoot for all simulations in the grid.  High-Z simulations have larger velocities which result in more convective overshoot.}
\label{fig:overshoottrends}
\end{figure}

Figure \ref{fig:overshoottrends} shows the size of the overshoot region for the entire set of simulations.  There appears to be a small correlation of overshoot with effective temperature, particularly in the high-Z simulations, but the dominant effect is clearly by varying metallicity. The larger degree of ‘overshoot’ in high-Z simulations is attributed to the larger mean and turbulent velocities.  The faster moving high-Z upflows carry more momentum and are able to traverse more space before losing their identity.

\section{Results: {\ttau} Relations} \label{sec:ttau}

Convection also affects the structure above the region of driven convection because of convective overshoot.  By directly solving the hydrodynamics equations, simulations provide a self-consistent representation of the effect of convection, including overshoot, on the atmosphere.  The primary way in which the stratification in and above the convective envelope is altered is from the additional support provided by turbulent pressure (described in Section \ref{sec:pturb}).  

1D Stellar structure models set the outer boundary condition by imposing a structure defined by a {\ttau} relation.  The Eddington {\ttau} is often used, but semi-empirical relations, such as those provided by \citet{1966ApJ...145..174K} and \citet{1981ApJS...45..635V} (VAL) can also be applied.  The semi-empirical relations implicitly include effects such as turbulent pressure, but are only valid for the Sun.  The location of the {\teff} surface with respect to optical depth is defined by the {\ttau} relation. The Eddington {\ttau} defines the {\teff} surface to be at $\tau=2/3$, while the semi-empirical relations show the {\teff} surface at smaller optical depths. Examining {\ttau} relations from simulations demonstrates how convective turbulence and overshoot affects the atmospheric structure, and to what degree it is sensitive to changes in metallicity.  

\begin{figure}[h]
\epsscale{1.2}
\plotone{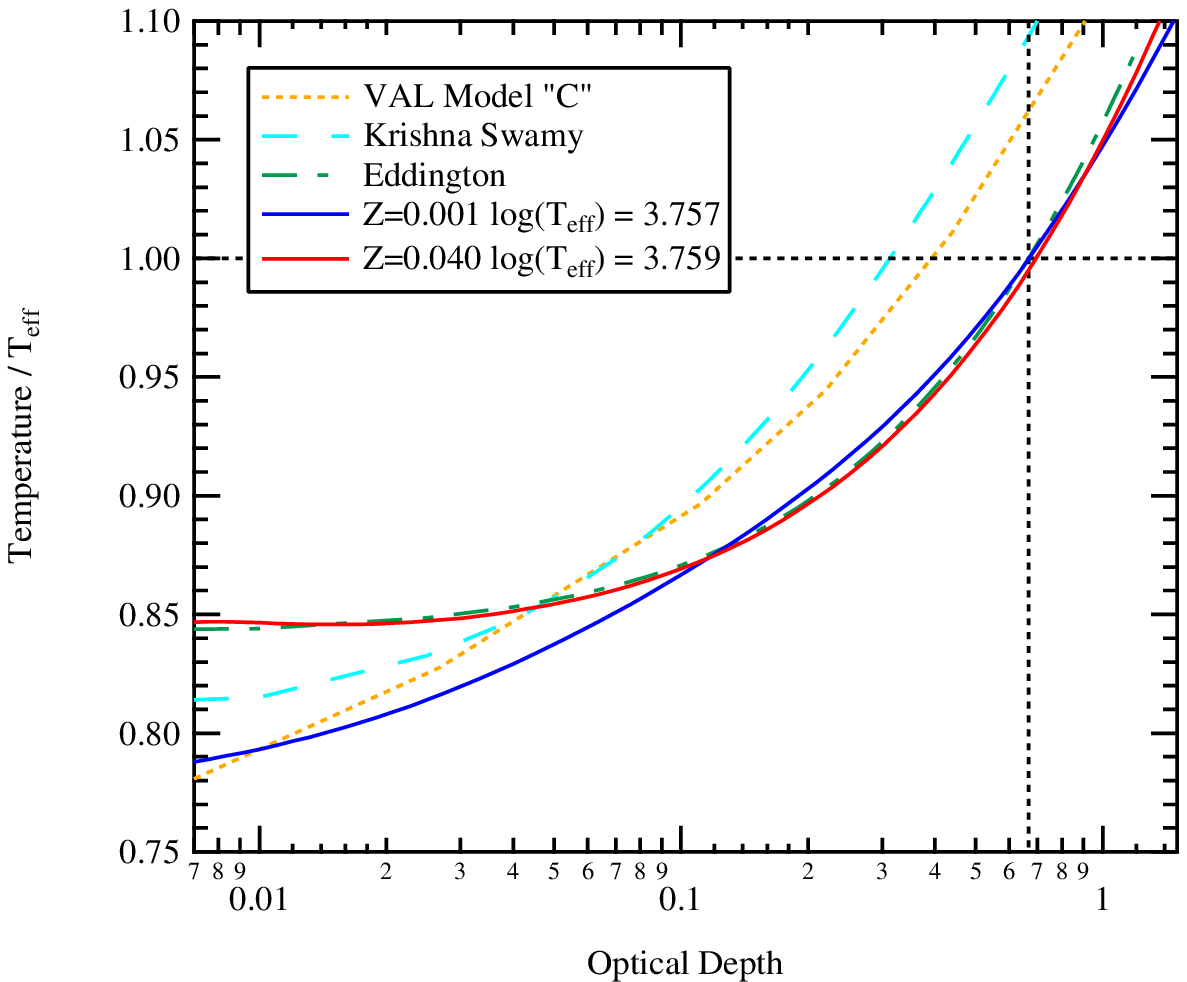}
\caption{Average {\ttau} relations for simulations s4 and s13 with varied metallicity, compared with analytic and semi-empirical {\ttau} relations for the Sun.  Low metallicity simulations have less radiative heating, resulting in a steeper {\ttau} relation in the optically thin layers.  Dotted lines mark the {\teff} surface and the $\tau=2/3$ optical surface, which by definition, occur at the same depth in the Eddington {\ttau} relation.}
\label{fig:ttau}
\end{figure}

The variation in {\ttau} relations with energy flux is rather small over the range of effective temperatures here.  In simulation sets of a given metallicity, the temperature variation at a given optical depth is typically less than a few percent.  The {\ttau} relation is more sensitive to changes in metallicity, however, as demonstrated in Figure \ref{fig:ttau}, which  shows {\ttau} relations for two simulations with the effective temperature but different metallicities.  The Eddington, Krisnha Swamy, and VAL {\ttau} relations are included for reference.  It is immediately apparent that the effect convective overshoot is sensitive to metallicity.  The comparison shows a discrepancy in the temperature at a given optical depth of up to 6\%.  

There is a clear systematic effect in response to the changing metalicities. Low-Z stratifications are cooler than high-Z stratifications at a given optical depth.  The example presented in Figure \ref{fig:ttau} shows that the high-Z and low-Z atmospheres have temperatures of approximately $T/T_{\rm eff}= 0.846$ and $0.793$ at $\tau=0.01$, respectively. 

Cooler atmospheric temperatures at low-Z in 3D simulations have been seen by \citet{1999A&A...346L..17A} and \citet{2011A&A...528A..32C}. They explain that the thermal structure of the atmosphere is determined by balancing the competing effects of radiative heating and adiabatic cooling from expanding upflows.  The decreased metallicity simulation has lower opacity and is less able to absorb radiation.  This leads to less radiative heating, so that balance is achieved at lower temperatures.

\section{Discussion}

\rev{Simulations of near-surface convection provide a realistic description of convection, and self-consistently couple the atmosphere and superadiabatic layer to the adiabatic region below.  The resulting stratification differs from, and is superior to 1D stellar models computed with MLT treatment of convection.  Differences between simulations and 1D models arise because the MLT treatment of convection lacks a description for some physical processes that are present in the simulations.  In particular, MLT does not include the turbulent contribution to pressure support.  Additionally, the upflowing granulation near the surface modifies the {\ttau} relation as a result of enhanced adiabatic cooling that 1D models do not include.  Providing a representation of processes like these in 1D stellar models could potentially improve their accuracy.}  

In Sections \ref{sec:salstructure} and \ref{sec:dynamics} we have shown how metallicity affects  near-surface convection in stars.   We are able to isolate the effect of metallicity by using a grid of simulations comprising sets of simulations at a fixed Z, but with overlapping {\teff} ranges.  Increasing metallicity increases the opacity, which alters the way in which energy transport transitions from convective to radiative.  

When the effective temperature is held fixed, changing the metallicity alters the mean stratification through the superadiabaitc region, with low-Z simulations exhibiting higher densities and pressures through the SAL.  \rev{Additional changes in pressure come from the contribution to hydrostatic equilibrium from turbulent pressure.  Because of the larger gas pressure at low-Z, the turbulent pressure relative to the total pressure becomes smaller, although it is still significant.}  The increase in density is required to compensate for the drop in opacity, in order to maintain the radiative flux.  This effect is expected, and certainly not specific to 3D simulations, but it has ramifications for other properties of convection present in the simulations.

The simulations show an adjustment of the superadiabatic excess as a result of changes to the mean stratification.  In the deep part of the convection zone, energy is transported almost entirely by convection and the temperature gradient is essentially adiabatic.  Nearer to the surface, the temperature gradient becomes superadiabatic as energy is transferred outward more efficiently by radiation.  The degree of superadiabatic excess marks the transition from convective to radiative energy transport, and is sensitive to changes in metallicity.  

Varying metallicity in the simulations induces remarkably little change in the maximum value of the superadiabaticity, even when the metallicity is reduced by a factor of 40.  \rev{The extent of the superadiabatic region also shows little remarkable variation between simulations.} This suggests that the rate of transition from convective to radiative energy transport is not drastically different in simulations of different metallicities.  The location of the transition, however, is quite different, as evidenced by a shift in the location of the superadiabatic peak. \rev{The SALs in the low-Z simulations occur at higher density.  The low-Z simulations have smaller opacity at a given density, and as a result the start of the transition to radiative energy transport occurs at higher density.}

The structure at small optical depth is set by the balance of radiative heating and adiabatic cooling from the convective updrafts.  3D simulations are able to account for accurate adiabatic cooling from the asymmetric flow that his characteristic of granulation.  Lower opacity in low-Z simulations results in less radiative cooling, and causes steeper {\ttau} gradients and cooler atmospheres.  This effect has been previously demonstrated using 3D simulations by \citet{1999A&A...346L..17A} and \citet{2007A&A...469..687C,2011A&A...528A..32C}. When metallicity is changed, our simulations show a substantial deviation (Figure \ref{fig:ttau} shows a change in temperature of up  to 8\%)  from the simple {\ttau} relations commonly used in stellar modeling.

Convective velocities are determined by the stratification, and the amount of energy flux that is transported.  Simulations with higher effective temperatures must transport more energy flux, and the convective velocities are expectedly larger.  The correlation of velocity and {\teff} exhibited in our simulation is consistent with the trend reported by \citet{2011ApJ...731...78T}. The energy fluxes are the same in simulations with the same {\teff} (but different Z), so differences in the convective velocities is a result of differences in the stratifications.  Because the high-Z simulations have lower density, the convective velocities must increase to maintain the energy flux.  Turbulent velocities are also strongly correlated with metallicity.  Our simulations show that the turbulent pressure \rev{as a fraction of gas pressure}, which is determined by the density and turbulent (rms) velocities, increases with effective temperature and metallicity.

Most of the simulation domain is convectively unstable, and, furthermore upflowing plumes do not simply stop in the radiative layers.  Instead, the momentum of the upflows carries them a certain `overshoot' distance before cycling back into the convection zone.  An expected consequence of larger mean and rms velocities is an increase in the overshoot distance.  We are able to measure boundaries on the overshoot region using correlations between turbulent fluctuations in temperature and velocity.  We find that high-Z simulations show significantly more overshoot.  Our simulations show a consistent and clear correlation between metallicity and overshoot, and in the most extreme case (between simulations with $Z=0.040$ and $Z=0.010$) we find a factor of $4.5$ change in overshoot distance.

The effect of convection, and its variation with stellar evolution, is typically not accurately represented in stellar modeling.  Most stellar models are calculated using MLT, or a similar prescription for convection, but there have been several attempts at including realistic convection in stellar models.  For example, \citet{1997ApJ...474..790D} introduced a variable mixing length parameter derived from the simulations of \citet{1996ApJ...461..499K}.  \citet{rosenthal1999} matched an envelope model to a solar simulation to include the effect of turbulent convection on oscillation frequencies, while \citet{li2002} included a two-parameter model for turbulent pressure and kinetic energy (calibrated with simulations) to 1D stellar models.  

The goal of simulating stellar envelope convection remains a generalized improvement over the current MLT-like treatments.    Our simulations show that that imposing a solar-calibrated mixing length parameter on all stellar models is not accurate, and ultimately, any generalized improvement for the treatment of convection in stellar models will have to account for composition.   \rev{Simulations can help improve stellar models by providing a self-consistent and realistic description of convection across the relative parameter space, which includes composition.  While metallicity is an important contribution to the composition, other significant factors, such as helium abundance, warrant consideration as well.  To that end, we are currently using simulations to investigate the effect of helium on convection.}

\acknowledgments

JT is supported by NASA ATFP grant \#NNX09AJ53G to SB, and acknowledges a PGS-D scholarship from the Natural Sciences and Engineering Research Council of Canada. This work was supported in part by the facilities and staff of the Yale University Faculty of Arts and Sciences High Performance Computing Center.

\end{document}